# Synergistic Formation of Radicals by Irradiation with both Vacuum Ultraviolet and Atomic Hydrogen: a Real-time *in situ* Electron Spin Resonance Study

*Kenji Ishikawa,[†][*] Naoya Sumi[†] Akihiko Kono,[‡] Hideo Horibe,[‡] Keigo Takeda,[†] Hiroki Kondo,[†] Makoto Sekine,[†] and Masaru Hori[†]*

[†]Nagoya University, Furo-cho, Chikusa-ku, Nagoya 464-8603, Japan

[‡]Kanazawa Institute of Technology, 3-1 Yatsukaho, Hakusan, Ishikawa 924-0838, Japan

Corresponding author; E-mail: ishikawa.kenji@nagoya-u.jp. Phone/Fax: +81 52 7886077.

ABSTRACT: We report on the surface modification of polytetrafluoroethylene (PTFE) as an example of soft- and bio-materials that occur under plasma discharge by kinetics analysis of radical formation using *in situ* real-time electron spin resonance (ESR) measurements. During irradiation with hydrogen plasma, simultaneous measurements of the gas-phase ESR signals of atomic hydrogen and the carbon dangling bond (C-DB) on PTFE were performed. Dynamic changes of the C-DB density were observed in real time, where the rate of density change was accelerated during initial irradiation and then became constant over time. It is noteworthy that C-DBs were formed synergistically by irradiation with both vacuum ultraviolet (VUV) and atomic hydrogen. The *in situ* real-time ESR technique is useful to elucidate synergistic roles during plasma surface modification.

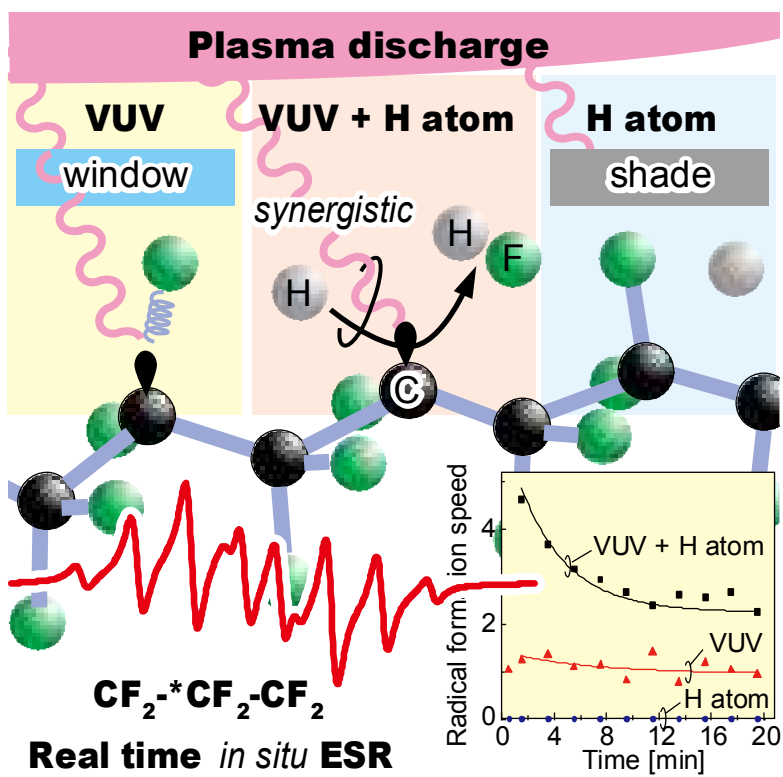

Light radiation can be considered as an ideal reagent for environmentally friendly, green chemical reactions, but poor absorption by organic substances makes direct solar photochemistry generally inefficient. Nevertheless, the need for high-energy ultraviolet (UV) radiation in most organic photochemical processes has limited both the practicality and environmental benefits.[1] For the development of green chemical processes, electrical discharge, known as cold plasma, rather than hot plasma, is beneficial for nuclear fusion reactors on both the laboratory and industrial scales. Many applications that utilize plasma, such as plasma-enhanced chemical vapor deposition, plasma etching, plasma-induced surface modification, and plasma medicine, have been widely recognized.[2,3] However, chemical reactions under plasma are very complicated, due to the simultaneous irradiation of electrons, ions, radicals, and photons. Therefore, the individual contributions of each of these reactive species in the plasma must be elucidated.

To understand the dynamic changes in bonding configurations during processing and surface reactions, direct observation of the formation and destruction of dangling bonds will provide much more important information. Originally, Westenberg reported that chemical reactions could be analyzed using the electron spin resonance (ESR) technique.[4,5] Previously, Yamasaki *et al.* succeeded in attempts to measure dangling bonds *in situ* during silicon film growth using ESR.[6] Subsequently, we have investigated the plasma processes for plasma etching of dielectric materials and diamonds using *in vacuo* and *in line* ESR techniques.[7-13]

Recently, the surface treatment of soft- and bio-material using atmospheric pressure plasmas has attracted much attention. For example, there is considerable interest in changing the surface hydrophobicity of polytetrafluoroethylene (PTFE) using treatments such as UV-radiation under vacuum at room temperature, which was studied by Rasoul *et al.*,[14] or oxygen low-pressure plasma exposure examined by Vandencasteele *et al.*,[15,16] Carbone *et al.*,[17] and Milella *et al.*[18] In addition, surface modification using atmospheric pressure He plasma was studied by Zettsu *et al.*[19] Physical and chemical modification of the PTFE surface has also reported by other researchers.[20-24]

Not only photochemical processes with plasma-induced emission, but also synergistic photo and radical processes play important roles in the utilization of plasma. Yuan *et al.* very recently reported that photochemical reactions on poly(methyl methacrylate) occurred synergistically from exposure to vacuum ultraviolet (VUV) and atomic oxygen.[25] In this study, *in situ* ESR measurements were performed during plasma processes using an ESR cavity and remote plasma system. PTFE has carbon-fluorine bonds with a relatively high bond energy of approximately 5 eV; therefore, it is difficult for photochemical oxidation to occur. Using plasma discharge, C dangling bond signals from the film were successfully observed during and after surface modification. However, determination of the dominant pathways under given conditions is challenging. ESR signals from both the solid and gas phases are presented with a tentative interpretation of their dynamic changes. These results have led to the elucidation of a plasma-induced synergistic process.

An ESR system was connected to a plasma discharge system using a quartz tube with an inner diameter of approximately 9 mm, as shown in Figure 1. A microwave (2.45 GHz) power supply of 50 W from a generator (Horonix MR-202) was used to generate plasma inside the quartz tube. Hydrogen gas was flowed into the quartz tube at a flow rate of 50 sccm and the pressure was maintained at approximately 10 Pa in the down-flow region. It should be noted that only hydrogen plasmas were used in this study.

ESR measurements were conducted using a standard X-band (9 GHz) spectrometer (Bruker Biospin, EMX plus) with a microwave resonator (Bruker Biospin, ER4119HS-W1). ESR spectra were recorded with a microwave power of 2 mW, a field modulation amplitude of 0.2 mT, and a modulation frequency of 100 kHz. All experiments were performed at room temperature. The quartz tube and PTFE film sample were inserted inside the ESR cavity in the down-flow region, typically 20 cm from the plasma discharge. The sample used was a 0.1 mm thick PTFE film (Fulon Industries). Both gaseous and surface radicals on the PTFE film were simultaneously detected *in situ*.

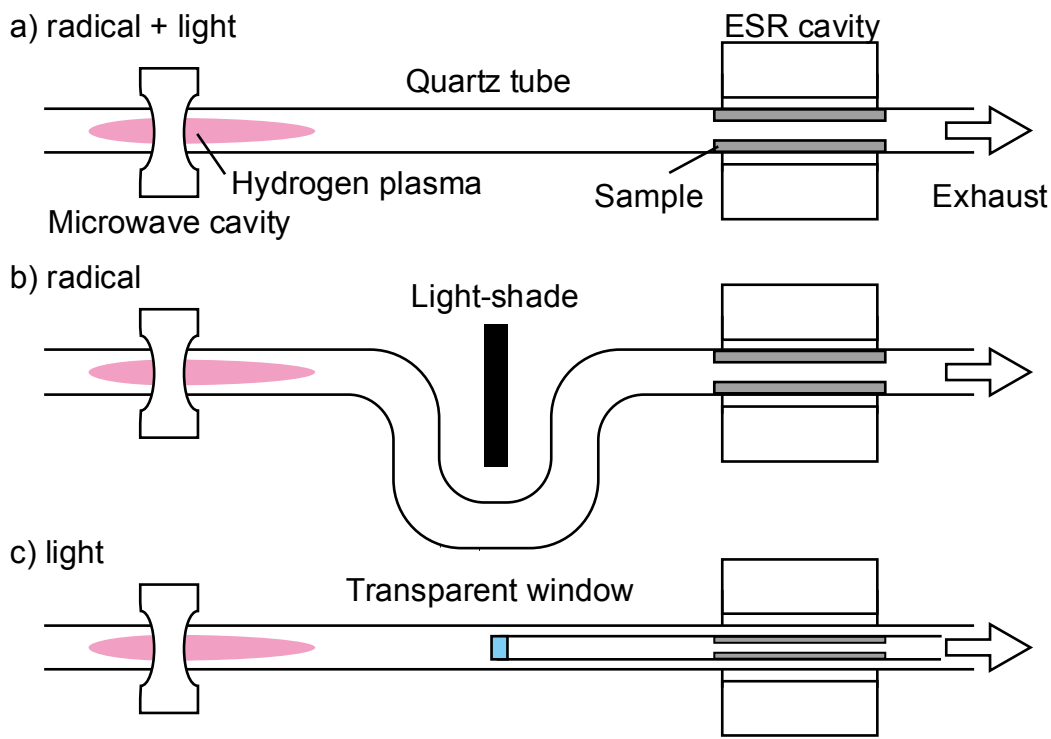

**Figure 1** Schematics of the experimental set-up. (a) Irradiation for plasma exposure with (b) only radicals and (c) only light irradiation.

Individual contributions from atomic hydrogen and plasma emission light have been studied in a similar manner using pallets for plasma process evaluation (PAPE).[26,27] The distance from the discharge was sufficiently large, so that charged particles such as electrons and ions were immediately diminished by recombination. Therefore, no effect of ion bombardment was observed at the down-flow region in this study. To limit interactions of plasma emissions with the sample, a U-shaped quartz tube and light shade were used, as shown in Fig. 1(b). Neutral species such as atoms and radicals are relatively long-lived; therefore, a sufficient amount of neutral species were transported to interact. In addition, interactions with only plasma emissions were clarified using the transparent $MgF_2$ window shown in Fig. 1(c), which can transmit wavelengths above approximately 140 nm.

Figure 2 shows experimental ESR signals during plasma discharge. Prior to plasma irradiation, only signals due to the quartz tube were observed in the ESR spectra, as shown in Fig. 2(a). During atomic H irradiation, a doublet ESR signal (g = 2.0114) for gaseous atomic H ($^2S_{1/2}$, I = 1/2) splitting of approximately 50.8 mT was observed.[6,28] Signals from atomic hydrogen are easily saturated at a microwave power of 2 mW; therefore, the spectrum shown in Fig. 2(a) was recorded at 0.02 mW. The ESR signals for H splitting indicated that atomic H was transported to the down-flow region at the ESR cavity.

Sharp ESR signals on the spectrum were detected, as shown in Fig. 2(b), only when the samples were inserted into the ESR cavity. Analysis of the spectra indicated carbon dangling bonds (C-DB) on the PTFE film. The signals correspond well with that simulated, assuming C-DBs (g = 2.0030) with surrounding fluorine atoms (I = 1/2, hyperfine interaction 9.1 mT for α-F and 3.4 mT for β-F) on PTFE, which has a fluoroalkyl chain structure of $CF_2$. The results indicate that the C-DB has the structure of $CF_2$-*CF-$CF_2$. In this analysis, other hyperfine splitting from $^{13}C$ (I = 1/2) was ignored because its natural abundance of 1.11% is relatively small.

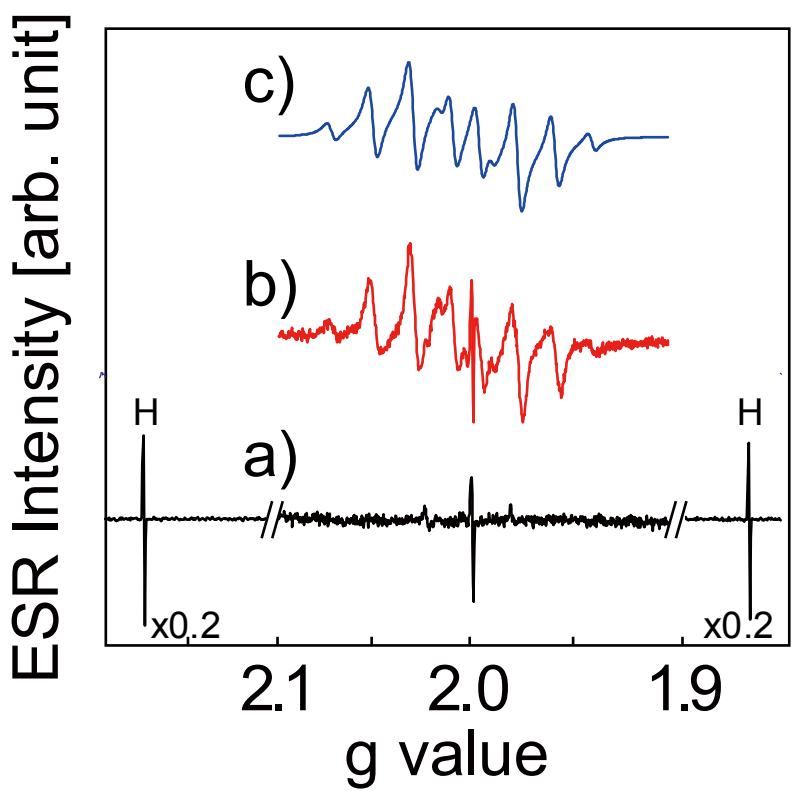

**Figure 2** ESR spectra of PTFE film; (a) before (initial quartz glass signal), and hyperfine doublet

(labeled "H") for atomic hydrogen (g = 2.0114, hyperfine interaction of 50.8 mT) during the plasma discharge; (b) after irradiation. (c) Simulated spectrum of the $CF_2$-*CF-$CF_2$ structure (g = 2.0030, I = 1/2, hyperfine interaction 9.1 mT for α-F and 3.4 mT for β-F).

The formation speed of surface radicals was measured and the roles of both atomic hydrogen and UV were analyzed using real-time measurements. The individual contributions from atomic hydrogen and plasma emission light were obtained. The difference in the rate of formation of these surface radicals for irradiation with atomic hydrogen, UV, and atomic hydrogen+UV is shown in Figure 3. When the sample was irradiated with VUV light, the C-DB signal was evident, which indicated that surface radicals were at least created by irradiation with VUV light. However, no C-DB was formed by irradiation with only atomic hydrogen.

The lack of C-DB formation by irradiation with only radicals is due to the relatively high bond-dissociation energy for C-F of approximately 5 eV. For VUV irradiation, a strong absorption energy of approximately 7.8 eV, which corresponds to a VUV wavelength of 160 nm, was reported by Nagayama *et al.*[29] This is in agreement with the C-DB created on the PTFE film by VUV irradiation in our experiments. (In addition, there was no formation of C-DB when the $MgF_2$ window material was changed to an optical filter that cut the wavelengths to as low as 250 nm.)

Notably, for irradiation with both atomic H and VUV, significant acceleration in the rate of C-DB formation was observed at the beginning of irradiation. If the reaction for scission of the fluoroalkyl chains is dominated by the flux of VUV photons, then the rate of formation of C-DB should remain constant because it is a first-order reaction. (Yuan *et al.* reported exponentially decreased reaction rates with decreasing sample thickness to that sufficiently thin for optical absorption.[25]) We emphasize that the rates were stabilized after a transient period of exposure time. After release of a fluorine atom, the C-DB formed would again be terminated by the released fluorine atom. In this situation, atomic hydrogen may play a scavenging role in reactions with atomic fluorine. The rate for the formation of dangling-bonds [db] is limited by the scavenger, which in this case is atomic hydrogen, [H]. This process is represented by

$$\frac{d[db]}{dt} \propto k(1-[db])[H], \quad [db] \propto (1-e^{-k[H]t}),$$

where $k$ is the rate coefficient. From this, acceleration in the rate as observed in the early period indicates that the synergetic formation of surface C-DB occurred close to the surface, due to the scission of C-F bonds, and was accelerated by VUV illumination and ruptured F atom scavenging by atomic H to form HF.

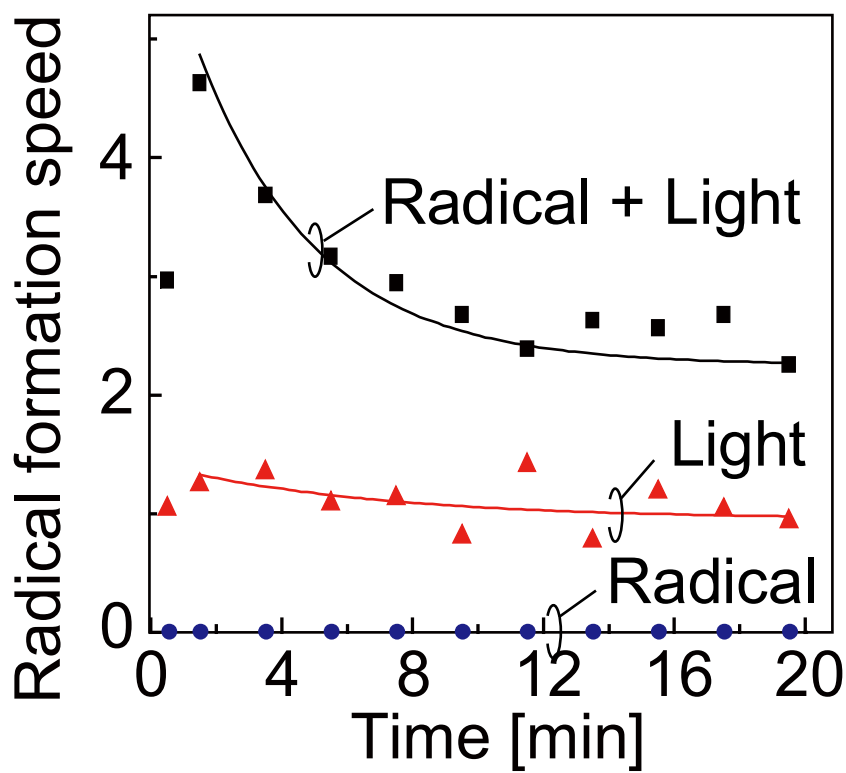

**Figure 3** Formation rate of C-DB determined from the ESR signal as a function of irradiation time.

Moreover, the surface radicals produced immediately changed to peroxy-radicals when the treated PTFE films were exposed to air.[24] This was confirmed by the introduction of oxygen gas, where the C-DB created on the PTFE film by atomic H and VUV were completely changed to peroxy-radicals, as shown in Figure 4. This suggests high reactivity of C-DB with oxygen and the peroxy-radical species can contribute to enhancement of the surface biocompatibility.

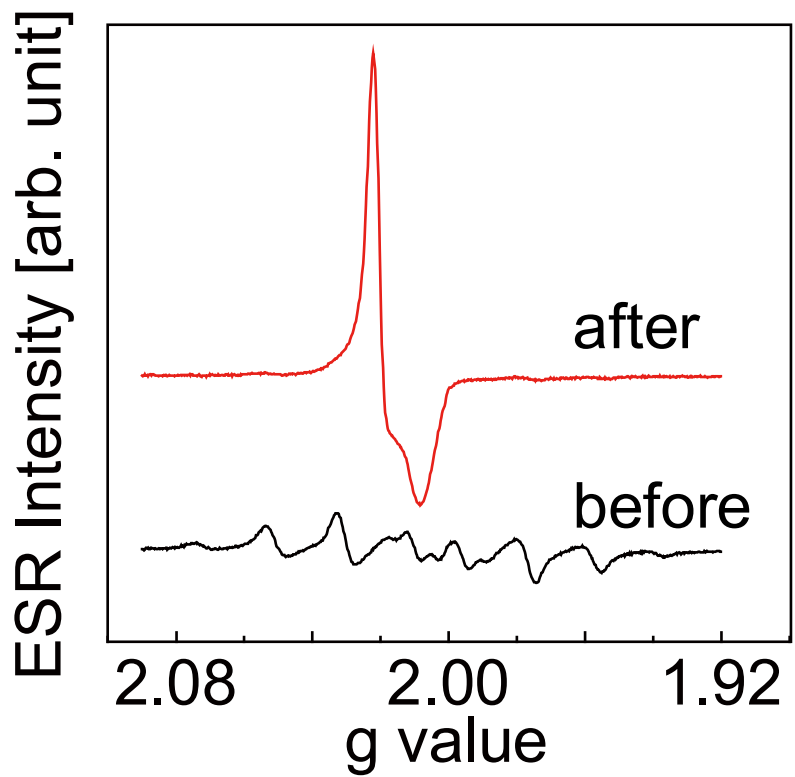

**Figure 4** ESR spectra of PTFE film showing peroxy-radical (-FCOO*, $g_\perp$ = 2.022 $g_{//}$ = 2.006) formation after oxygen exposure of a sample (identical to Fig. 2) that has C-DBs on fluoroalkyl-chains by the formation of hydrogen atoms and VUV light irradiation.

The *in situ* measurement system is an effective approach to determine radical formation interactions between surface F atoms and gaseous H atoms. This study revealed the synergistic and photocatalytic rupture of C-F bonds by VUV light irradiation and inhibition of recombination by the formation of HF due to the supply of atomic hydrogen.

In summary, C-DBs created by plasma discharge were studied using *in situ* real-time ESR

measurements. The PTFE surface was used as an example of a soft material and was exposed to H atoms. During atomic H exposure, the ESR spectra exhibited C-DB signals. The *in situ* real-time ESR technique was demonstrated as a new experimental approach to the microscopic understanding of chemical reactions on surfaces with gaseous radicals during plasma processes. We have successfully obtained information regarding the reaction mechanism with radicals generated by plasma induced surface interactions.

ACKNOWLEDGMENT

This work was supported in part by the Knowledge Cluster Initiative (Second Stage) of the Tokai Region Nanotechnology Manufacturing Cluster.

REFERENCES

(1) Yoon, T. P.; Ischay, M. A.; Du, J. Visible Light Photocatalysis as a Greener Approach to Photochemical Synthesis, *Nature Chem.* **2010**, *2*, 527-532.

(2) Lieberman, M. A.; Litchtenberg, A. J. *Principles of Plasma Discharges and Materials Processing*, Wiley-Interscience, New York, **2005**.

(3) Fridman, A. *Plasma Chemistry*, Cambridge University Press, New York, **2008**.

(4) Westenberg, A. A. Quantitative Measurements of Gas Phase O and N Atom Concentrations by ESR, *J. Chem. Phys.* **1964**, *40*, 3087-3098.

(5) Westenberg, A. A. Applications of Electron Spin Resonance to Gas-Phase Kinetics, *Science* **1969**, *164*, 381-388.

(6) Yamasaki, S.; Umeda, T.; Isoya, J.; Tanaka, K. In Situ Electron-Spin-Resonance Measurements of Film Growth of Hydrogenated Amorphous Silicon, *Appl. Phys. Lett.* **1997**, *70*, 1137-1139.

(7) Ishikawa, K.; Kobayashi, S.; Okigawa, M.; Sekine, M.; Yamasaki, S.; Yasuda, T.; Isoya, J. In Vacuo Electron-Spin-Resonance Study on Amorphous Fluorinated Carbon Films for Understanding of Surface Chemical Reactions in Plasma Etching, *Appl. Phys. Lett.* **2002**, *81*, 1773-1175.

(8) Ishikawa, K.; Okigawa, M.; Ishikawa, Y.; Samukawa, S.; Yamasaki, S. In Vacuo Measurements of Dangling Bonds Created during Ar-diluted Fluorocarbon Plasma Etching of Silicon Dioxide Films, *Appl. Phys. Lett.* **2005**, *86*, 264104:1-3.

(9) Ishikawa, K.; Yamaoka, Y.; Nakamura, M.; Samukawa, S.; Ishikawa, Y.; Yamazaki, Y.; Yamasaki, S. Surface Reactions during Etching of Organic Low-k Films by Plasmas of N and H, *J. Appl. Phys.* **2006**, *99*, 083305:1-6.

(10) Yamazaki, Y.; Ishikawa, K.; Mizuochi, N.; Yamasaki, S. Etching Damage in Diamond Studied using an Energy-Controlled Oxygen Ion Beam, *Jpn. J. Appl. Phys.* **2007**, *46*, 60-64.

(11) Yamazaki, Y.; Ishikawa, K.; Samukawa, S.; Yamasaki, S. Defect Creation in Diamond by Hydrogen Plasma Treatment at Room Temperature, *Physica B* **2006**, *376/377*, 327-330.

(12) Yamazaki, Y.; Ishikawa, K.; Mizuochi, N.; Yamasaki, S. Structural Change in Diamond by Hydrogen Plasma Treatment at Room Temperature, *Diamond. Related Mater.* **2005**, *14*, 1939-1942.

(13) Yamazaki, Y.; Ishikawa, K.; Mizuochi, N.; Yamasaki, S. Structure of Diamond Surface Defective Layer Damaged by Hydrogen Ion Beam Exposure, *Diamond. Related Mater.* **2006**, *15*, 703-706.

(14) Rasoul, F. A.; Hill, D. J. T.; George, G. A.; O'Donnell, J. H. A Study of a Simulated Low Earth Environment on the Degradation of FEP Polymer, *Polym. Adv. Technol.* **1998**, *9*, 24-30.

(15) Vandencasteele, N.; Nisol, B.; Viville, P.; Lazzaroni, R.; Castner, D. G.; Reniers, F. Plasma-Modified PTFE for Biological Applications: Correlation between Protein Resistant Properties and Surface Characteristics, *Plasma Process Polym.* **2008**, *5*, 661-671.

(16) Vandencasteele, N.; Broze, B.; Collette, S.; De Vos, C.; Viville, P.; Lazzaroni, R.; Reniers, F. Evidence of the Synergetic Role of Charged Species and Atomic Oxygen in the Molecular Etching of PTFE Surfaces for Hydrophobic Surface Synthesis, *Langmuir* **2010**, *26*, 16503-16509.

(17) Carbone, E. A. D.; Boucher, N.; Sferrazza, M.; Reniers, F. How to Increase the Hydrophobicity of PTFE Surfaces Using an RF Atmospheric-Pressure Plasma Torch, *Surf. Interf. Anal.* **2010**, *42*, 1014-1018.

(18) Milella, A.; Palumbo, F.; Favia, P.; Cicala, G.; d'Agostino, R. Continuous and Modulated Deposition of Fluorocarbon Films from *c*-$C_4F_8$ Plasmas, *Plasma Process. Polym.* **2004**, *1*, 164-170.

(19) Zettsu, N.; Itoh, H.; Yamamura, K. Surface Functionalization of PTFE Sheet through

Atmospheric Pressure Plasma Liquid Deposition Approach, *Surf. Coat Technol.* **2008**, *202*, 5284-5288.

(20) Clark, D. T.; Hutton, D. R. Surface Modification by Plasma Techniques. I. The Interactions of a Hydrogen Plasma with Fluoropolymer Surfaces, *J. Polym. Sci. A: Polym. Chem.* **1987**, *25*, 2643-2664.

(21) Kuzuya, M.; Ito, H.; Kondo, S.; Noda, N.; Noguchi, A. Electron Spin Resonance Study of the Special Features of Plasma-Induced Radicals and Their Corresponding Peroxy Radicals in Polytetrafluoroethylene, *Macromole.* **1991**, *24*, 6612-6617.

(22) Haupt, M.; Barz, J.; Oehr, C. Creation and Recombination of Free Radicals in Fluorocarbon Plasma Polymers: An Electron Spin Resonance Study, *Plama Process. Polym.* **2008**, *5*, 33-43.

(23) Oshima, A.; Seguchi, T.; Tabata, Y. ESR Study on Free Radicals Trapped in Crosslinked Polytetrafluoroethylene (PTFE), *Radiat. Phys. Chem.* **1997**, *50*, 601-606.

(24) Oshima, A.; Seguchi, T.; Tabata, Y. ESR Study on Free Radicals Trapped in Crosslinked Polytetrafluoroethylene (PTFE)-II Radical Formation and Reactivity, *Radiat. Phys. Chem.* **1999**, *55*, 61-71.

(25) Yuan, H.; Killelea, D. R.; Tepavcevic, S.; Kelber, S. I.; Sibener, S. J. Interfacial Chemistry of Poly(methyl methacrylate) Arising from Exposure to Vacuum-Ultraviolet Light and Atomic Oxygen, *J. Phys. Chem. A* **2010**, *115*, 3736-3745.

(26) Uchida. S.; Takashima, S.; Hori, M.; Fukasawa, M.; Ohshima, K.; Nagahata, K.; Tatsumi, T. Plasma Damage Mechanisms for Low-k Porous SiOCH Films due to Radiation, Radicals, and Ions in the Plasma Etching Process, *J. Appl. Phys.* **2008**, *103*, 073303:1-5.

(27) Moon, C. S.; Takeda, K.; Sekine, M.; Setsuhara, Y.; Shiratani, M.; Hori, M. Etching Characteristics of Organic Low-k Films Interpreted by Internal Parameters Employing a Combinatorial Plasma Process in an Inductively Coupled H/N Plasma, *J. Appl. Phys.* **2010**, *107*, 113310:1-8.

(28) Murata, T.; Shiraishi, K.; Ebina, Y.; Miki, T. An ESR Study of Defects in Irradiated Hydroxyapatite, *Appl. Radiat. Isot.* **1996**, *47*, 1527-1531.

(29) Nagayama, K.; Miyamae, T.; Mitsumoto, R.; Ishii, H.; Ouchi, Y.; Seki, K. Polarized VUV Absorption and Reflection Spectra of Oriented Films of Poly(tetrafluoroethylene)$(CF_2)_n$ and its Model Compound, *J. Electron. Spectrosc. Related Phenomena* **1996**, *78*, 407-410.